\documentclass[a4paper,fleqn,usenatbib,useAMS]{mnras}

\usepackage{graphicx}	
\usepackage{amsmath}	
\usepackage{amssymb}	
\usepackage{multicol}        
\usepackage{bm}		
\usepackage{pdflscape}	

\usepackage[T1]{fontenc}
\usepackage{ae,aecompl}

\usepackage{newtxtext,newtxmath}

\title[MWBs as tracers of massive star formation]{Massive, wide binaries as tracers of massive star formation}

\author[D.~W.~Griffiths et al.]{Daniel~W.~Griffiths$^{1}$, Simon~P.~Goodwin$^{1}$ and Saida~M.~Caballero-Nieves$^{2,1}$
\\
$^1$ Dept. Physics \& Astronomy, University of Sheffield, Hicks Building, Hounsfield Road, Sheffield, S3 7RH, UK\\
$^2$ Department of Physics and Space Sciences, Florida Institute of Technology, 150 W. University Boulevard, Melbourne, FL 32901, USA }

\pubyear{2018}

\begin{document}
\label{firstpage}
\pagerange{\pageref{firstpage}--\pageref{lastpage}}
\maketitle

\begin{abstract}
Massive stars can be found in wide (hundreds to thousands~AU) binaries with other massive stars.  We use $N$-body simulations to show that any bound cluster should always have approximately one massive wide binary: one will probably form if none are present initially; and probably only one will survive if more than one are present initially.  Therefore any region that contains many massive wide binaries must have been composed of many individual subregions. Observations of Cyg~OB2 show that the massive wide binary fraction is at least a half (38/74) which suggests that Cyg~OB2 had at least 30 distinct massive star formation sites. This is further evidence that Cyg~OB2 has always been a large, low-density association.  That Cyg~OB2 has a normal high-mass IMF for its total mass suggests that however massive stars form they `randomly sample' the IMF (as the massive stars did not `know' about each other).
\end{abstract}




\section{Introduction}
\label{section: introduction}

\bigskip
\bigskip

How stars form is one of the key questions in astrophysics.  Of
particular importance is how the rare, but extremely influential, massive stars form.  

The most popular massive star formation models fall into two main
camps: `isolated' (e.g. \citealt{Krumholz2005}), and `competitive'
(e.g. \citealt{Bonnell1997}).  Massive star formation is reviewed in
detail by \citet{Zinnecker2007a}, but generally: in 
`isolated' formation massive stars form from very massive
cores and are `destined' to be massive; while in `competitive' models
initially low-mass stars `lucky' enough to be in regions of high gas
density can grow to become massive.

To some extent, the distinction between `isolated' and `competitive' models is if massive stars form in `clustered' environments or `associations'.  Here we use 
`cluster' to refer to bound groups of stars, and `associations' as unbound groups of stars.  In a clustered environment stars are expected to encounter one another and `know' that other stars are present, which is not necessarily true in an association.  (We have rather simplified the arguments here, see \citet{Zinnecker2007a} for a more in-depth discussion.)

Distinguishing between these models of massive star formation is
difficult.  A common prediction of competitive models is that massive
stars require a gas- and star-rich dynamical environment to form, and so massive
stars will form in `clusters', but in isolated models massive
stars can form in regions with few other stars with no `knowledge' of other star formation.  

This has motivated searches for `isolated' massive stars which are not associated with
`clusters' (e.g. \citealt{Lamb2010}; \citealt{Oey2013}; \citealt{Bressert2012}).
However, it is known that some/many isolated massive stars have been
ejected from dense clusters \citep{Fujii2011,Oh2015} and so a definitive
identification as a massive star as having {\em formed} in relative
isolation is difficult. 

Another approach is to examine the massive star population of associations.  For example, Cyg OB2 has a mass of $\sim 10^5$ M$_\odot$ and a full IMF of massive stars up to 100 M$_\odot$ (\citealt{Wright2015}).  With a size of $\sim 20$~pc,  and a velocity dispersion of $\sim 20$~km~s$^{-1}$ (Wright et al. 2016), Cyg OB2 has a virial ratio of $\sim 10$ and is a (highly) unbound association.  However, all we can say is that Cyg OB2 is unbound at its current age of 2--10 Myr (it has a significant internal age spread), but it is unclear if the regions in which the massive stars formed were `clustered' and have since expanded (although the structure of the association suggests not, Wright et al. 2014).

In this paper we investigate massive wide binaries (MWBs) as a signature of how massive stars form.  A MWB is two massive stars in a binary that is potentially wide enough to be dynamically destroyed or altered.  Because such binaries are
susceptible to destruction in dense 
environments, they can carry information on the density history of their
environment.

We define a MWB as a binary system in which both stars have masses 
greater than 5~M$_{\odot}$, and which have a separation, $a$, 
between $10^{2}<a<10^{4}$ AU.  There are three things that make such MWBs ($>$5~M$_{\odot}$) particularly interesting.

Firstly, because the primaries and secondaries are both bright (O, B or A-stars) and
well-separated they are relatively easy to find as visual binaries
even at large distances. Later we discuss the observed MWBs in Cyg~OB2, and in the observations of Caballero-Nieves et al. (in prep.; our choice of $\sim 5 M_\odot$ is partly motivated by the detection limit of this survey, but this is not very important to our results).

Secondly, even at such wide separations they are intermediate, or 
even hard binaries in that {\em low-mass} stars do not
carry enough energy to disrupt them, as 5~M$_{\odot}$ is significantly more massive than a `typical' star (0.2--0.5$M_\odot$; see below for more details). Therefore MWBs are only susceptible to disruption by other `massive' stars.

Thirdly, MWBs are the only type of binary system that can be easily
produced by three-body encounters between stars (again, see below). 

Therefore, the numbers of MWBs in a region should provide evidence of the past density and dynamical history of that region, in particular the past history of the massive stars.

\section{Binary formation and destruction}

We wish to investigate the different environments in
which massive wide binaries (MWBs) can survive, are destroyed, or can form (or some mixture of the three can occur).

\subsection{Binary formation}

The binary formation rate per
unit volume $\dot{N_{\rm b}}$, as a function of stellar mass $m$, stellar
number density $n$ and velocity dispersion $\sigma$, is given by
\cite{Hut1992} as: 
\begin{equation}
\dot{N_{\rm b}} = 0.75 \frac{G^{5}m^{5}n^{3}}{\sigma^{9}}
\label{equation: binaryFormation}
\end{equation}
While this rate is negligible for Galactic field stars, its dependence
on the density and velocity dispersion of a region means that it can be
significant for dense clusters \citep{Reipurth2014}. Furthermore, the
dependence of Eqn \ref{equation: binaryFormation} on the stellar
mass $m^5$ indicates that high-mass binaries will form at a much faster rate than 
their low-mass counterparts. 
\citet{Moeckel2011} find that soft binaries are continually created
as well as destroyed (e.g. \citet{Heggie1975}) dense environments, and \citet{Allison2011} show that
massive stars can form binaries that harden, and even form Trapezium-like systems,
which can survive in the long term.

From Eqn \ref{equation: binaryFormation} 
we expect MWB formation to depend on the (number) density of massive stars 
(the $n^{3}$ term), moderated by the velocity dispersion ($\sigma^{-9}$).  So we would expect more MWBs to form at higher densities and in the presence of other massive stars.  This is rather non-trivial as as higher densities usually mean higher velocity dispersions, so in a virialised cluster with radius $R$ we would expect $\sigma^9 \propto n^{9/2} R^9$, so $n^3/\sigma^9 \propto 1/(n^{3/2}R^9)\,$.

\subsection{Binary destruction}

A binary system can be categorised as either a `hard', `intermediate'
or `soft' binary according to the difference between the binding
energy of the binary $|E_{\rm bind}|$ and the typical energy in an 
encounter $E_{\rm enc}$ \citep{Hills1975,Heggie1975,Hills1990}.
When $|E_{\rm bind}| >> E_{\rm enc}$ the binary is `hard':
ie. encounters will be unable to destroy or significantly alter the
binary.
When $|E_{\rm bind}| << E_{\rm enc}$ the binary is `soft':
ie. encounters will very quickly destroy the binary.
When $|E_{\rm bind}| \sim E_{\rm enc}$ the binary is `intermediate': i.e. it
may survive or may be destroyed depending on the details of its
encounter history (see \citealt{Parker2012a}).

As shown by \citet{Hills1990} it is often better to consider the velocity of a perturber, rather than simply the energy.  During an encounter of a binary system with primary and secondary masses $m_{\rm p}$ and $m_{\rm s}$ and semi-major axis $a$, with a perturbing star with mass $m_{\rm pert}$, the critical velocity $v_{\rm c}$ is defined as the velocity at which the total energy of the three bodies involved in the encounter is zero, given by:
\begin{equation}
v_{\rm c} = \frac{G m_{\rm p} m_{\rm s} (m_{\rm p} + m_{\rm s} + m_{\rm pert})}{m_{\rm pert}(m_{\rm p} + m_{\rm s}) \,a}
\label{equation: criticalvelocity}
\end{equation}
If the perturber velocity $v_{\rm pert} << v_{\rm c}$, then the binary will not be destroyed. However the properties of the binary may be altered by an energy exchange, and it is possible to have an exchange of members (typically if the perturber is of higher mass than the secondary). From Eqn. \ref{equation: criticalvelocity} we can see that for a MWB comprised of two 5 M$_{\odot}$ stars, in order to destroy the binary, the velocity of a 1 M$_{\odot}$ perturber would need to be over three times larger than that of a 50 M$_{\odot}$ perturber.

Whether a binary will survive or be disrupted depends not only on the energy/velocity of an encounter, but the rate of encounters close enough to disrupt the binary.  The encounter rate, $t_{\rm enc}$, 
is inversely proportional to both the number density and velocity dispersion, $\propto 1/(n\sigma)$ (see e.g. \citep{Binney1987}).  In a virialised cluster  
of radius $R$, the encounter rate will therefore depend on the crossing (dynamical) timescale, 
$t_{\rm cross} = R/\sigma$, of the cluster as $t_{\rm enc} \propto t_{\rm cross}^3/R$.  In addition, the velocity of encounters has a dependency $\sigma \propto n^{1/2}R$ which complicates any estimates of encounter rates.

For MWBs the encounter rate has another subtlety.  The number density of interest is not the number density of all stars, but rather the number density of stars massive enough to potentially destroy the binary.  Generally this will be significantly lower than the `average' number density, but can be enhanced by (primordial or dynamical) mass segregation (which can then reduce the velocity dispersion of the massive stars so reducing their encounter energy).

There is yet another subtlety that needs to be borne in mind: encounters can harden a binary, in particular if the encounter velocity is $<< v_{\rm c}$ (the Heggie-Hills Law).  This can mean that a massive binary with an initial separation greater than the nominal 100~AU limit for `wide' can be hardened below this limit and `drop out' of a MWB sample (we see this effect later).  This depends on the encounter rate in the same way as destructive encounters, but if hardening or softening encounters dominate depends on the each encounter energy relative to the particular MWB energy.

The above discussion shows that the rate at which MWBs are destroyed is rather complex and has no simple dependencies on time-scales such as the crossing time.  The binary destruction rate will also be rather stochastic depending on if a MWB has an encounter with enough energy to destroy it (see e.g. \citealt{Parker2012a}), or if encounters harden a binary below a nominal limit. Ensembles of $N$-body simulations are required to investigate the interplay of all of these effects.

\section{Method/Initial Conditions}
\label{section: method}
We perform ensembles of $N$-body simulations using the {\sc KIRA}  
$N$-body integrator from the Starlab package \citep{PortegiesZwart2001}.

Throughout we define a MWB as a 
binary system comprised of two stars each with masses greater than 5
M$_{\odot}$, with an instantaneous 3D separation between $10^2$ and $10^4$ AU.  Note
that the instantaneous 3D separation is not the same as the semi-major
axis of the orbit (it is likely to be somewhat larger, depending on the
eccentricity of the orbit and the current phase), and it is not the
same as the projected separation that would be observed.  We pick
the instantaneous 3D separation for
simplicity due to the dependence of the projected separation on viewing angle (the instantaneous separation
is therefore an upper limit on any projected separation).

Every simulation starts as a virialised Plummer sphere
\citep{Plummer1911OnClusters} 
with a total mass of stars $M_{T}\approx600$ M$_{\sun}$ ($\sim 400$ stars $>0.1 M_\odot$).  We pick $\sim 600 M_\odot$ as that is the mass at which we would expect one or two O-stars ($>20 M_\odot$) if randomly sampling from a standard IMF.

The stars in each simulated region are allocated a position and
velocity using the method described in \citet{Aarseth1974}. The
timescale of each simulation is 10~Myr, and no stellar evolution is 
included.  

Whilst virialised Plummer spheres are very simple initial conditions,
we expect {\em any} initial distribution to relax to something similar to a virialised Plummer-like
distribution within a few initial crossing times as long as it is
initially bound (see e.g. \citealt{Allison2009}; \citealt{Allison2011}).

We perform two sets of simulations: set `N' that start with no MWBs,
and set `B' in which we place a `primordial' MWB\footnote{Although we
  note that this MWB could have formed dynamically during an earlier
  relaxation phase of the region which we ignore.}.

For all of the primordial binary `B' scenarios, the primordial MWB is
composed of two stars, star $\alpha$ and star $\beta$. Star
$\alpha$ is the primary star in the primordial binary, and has a mass
of 20 M$_{\odot}$. The secondary, star $\beta$, mass is uniformly randomly
sampled between 10 M$_{\odot}$ and 20 M$_{\odot}$, giving a 
binary mass ratio of $0.5\leq q_{\alpha\beta}\leq
1.0$. The binary separation for these primordial binaries is chosen
uniformly between 1000 and 5000 AU (within our working definition of a MWB), and the eccentricity is set to zero. 

For all of the `N' scenarios, stars $\alpha$ and $\beta$, which make up
the primordial binaries in the `B' scenarios, are still present. However, they
are not part of a binary system but are instead single stars,
randomly placed in the Plummer sphere. 

\bigskip

We run ensembles of 100 simulations in which we vary only the random
number seed used to set the initial conditions.  Each ensemble is run
with (B) and without (N) a primordial MWB in one of four scenarios
(see below) with four different initial densities (see below). 

{\bf 1. All other stars are low-mass.}  In ensembles N1 and B1 all stars other than $\alpha$ and $\beta$ (be
they part of a binary or two single stars) have a mass of 1~M$_\odot$.

\begin{table}
  \begin{tabular}{| l | l | l | l | l}
	\hline
	Scenario & Mass               & No. of $>30$        & Primordial & $t_{\rm cross}$ \\
 	         & Function           & M$_{\odot}$ Stars  & Binary? & Myr \\
    \hline
	N1       & Flat 1 M$_{\odot}$  & 0                 & No & 0.08 \\
	N2       & Maschberger        & 0                 & No & 0.25 \\
	N3       & Maschberger        & 1                 & No & 0.66 \\
	N4       & Maschberger        & 5                 & No & 1.2 \\
	\hline
	B1       & Flat 1 M$_{\odot}$  & 0                 & Yes & 0.08\\
	B2       & Maschberger        & 0                 & Yes & 0.25 \\
	B3       & Maschberger        & 1                 & Yes & 0.66 \\
	B4       & Maschberger        & 5                 & Yes & 1.2 \\
	\hline
  \end{tabular}
\caption{A summary of the differences in the initial conditions.  In the first column, scenarios are numbered 1--4 with `N' for no primordial MWBs, and `B' for an primordial MWB (repeated in column 4 for clarity).  The second column has the stellar mass function used (flat or `normal').  The third column has the number of very massive stars ($> 30 M_\odot$) in the cluster.}
\label{table:initialConditions}
\end{table}

{\bf 2. All other stars are lower-mass with a normal IMF.}  In ensembles N2 and B2, all of the stars which make up the cluster,
except for stars $\alpha$ and $\beta$, have masses randomly sampled
from the standard single star Maschberger IMF 
\citep{Maschberger2013OnFunction}. A lower limit of $0.1 M_{\sun}$
prevents the inclusion of brown dwarfs and other objects with masses
far too low to affect the binary, the upper limit of $10 M_{\sun}$
means that the stars $\alpha$ and $\beta$ are the most massive stars
in the cluster.  (We force the masses of the two most massive stars to be 10--20 and $20 M_\odot$, but as mentioned above this would be expected for this total cluster mass.)

{\bf 3. The cluster includes one more massive star.} Ensembles N3 and B3 are the same as N2 and B2 but with the addition of
a single new higher-mass star, with a mass between 30-35 M$_{\sun}$, to the
cluster.

{\bf 4. The cluster contains three more (single) massive stars.} The last ensembles, N4 and B4, add 3 more massive stars to the
cluster, each with masses between 30 and 50 M$_{\sun}$. 

Note that a higher mass limit on the background cluster stars of 10 M$_{\odot}$ allows for the existence of more stars with masses greater than 5 M$_{\odot}$, from which a MWB could form. In total, there are up to $\sim20$ stars with masses greater than 5 M$_{\odot}$ in each of the ensembles N2-N4 and B2-B4. In principle, any of these could form a MWB in our definition of a MWB.

In each of the eight
ensembles above, the clusters are given four different initial
densities: half-mass radii, $R_{1/2}$, of 0.25, 0.5, 1 and 1.5~pc.   For a cluster with a half-mass radius between $0.25\leq R_{1/2}\leq
1.5$ pc, the half-mass density (in M$_\odot$ pc$^{-3}$) is $1.25\leq
\log\rho_{1/2}\leq 3.58$, the upper limit of this is of a similar
density to the Arches cluster, while the lower limit is similar to RSGC03 \citep{PortegiesZwart2010} (both clusters contain several massive stars). 

For reference, the crossing times of the clusters are roughly 0.08, 0.25, 0.66 and 1.2 Myr for $R_{1/2}=$ 0.25, 0.5, 1.0 and 1.5 pc respectively.  Note that while it is possible to calculate a relaxation time for these clusters, that number is rather difficult to interpret or give any meaning too as $N$ is so low.

Table \ref{table:initialConditions} gives a summary of the different
initial conditions in each of the eight scenarios, N1-N4 and B1-B4,
based on the mass distribution of stars in the cluster, and whether
stars $\alpha$ and $\beta$ begin in a primordial massive wide binary
or whether they begin as single stars.

\section{Results}
\label{section: results}

We will first consider the formation of MWBs in ensembles that start
with no binaries (N1-N4), and then both the formation {\em and}
destruction of MWBs in ensembles with primordial MWBs (B1-B4).

\subsection{The formation of MWBs}
\label{section: resultsScenarioA}

\begin{table}
\centering
\begin{tabular}{ | l | l | l | l | l |}
  \hline
  Scenario & \multicolumn{4}{| c |}{Number of Simulations Containing a Massive,} \\
           & \multicolumn{4}{| c |}{Wide Binary at $t=10$ Myr} \\
  \hline
     & $R_{1/2}=0.25$ pc & 0.50 pc & 1.00 pc & 1.50 pc \\
  \hline
  N1 & 81                & 73      & 3        & 0       \\     
  N2 & 63                & 74      & 16       & 5       \\     
  N3 & 92                & 89      & 16       & 2       \\     
  N4 & 87                & 82      & 22       & 1       \\     
\end{tabular}
\caption{Number of MWBs which formed in clusters with different initial half-mass radii $R_{1/2}$, for each of the no primordial MWB Scenarios N1-N4.}
\label{table:tableAFormed}
\end{table}

\begin{figure}
\centering
\includegraphics[]{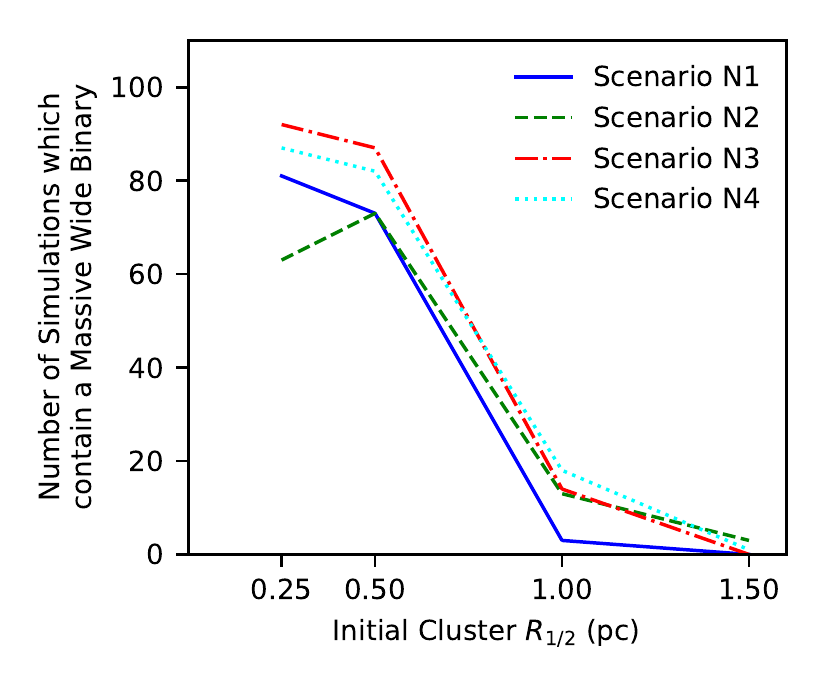}
\caption{Number, out of 100, of clusters with no primordial MWB which
  contain at least one MWB after 10~Myr, as a
  function of the initial cluster half-mass radius, $R_{1/2}$, for Scenarios N1 (blue solid line), N2 (green dashed line), N3 (red dot-dashed line), and N4 (cyan dotted line).
  \label{fig:ScenarioAFormed}} 
\end{figure} 

All simulations N1-N4 initially contain no binary systems.
Table~\ref{table:tableAFormed} shows the number (out of 100) of simulations
in which a MWB is found to be present after 10~Myr for each scenario (N1-N4)
at each density ($R_{1/2}=0.25, 0.5, 1, 1.5$~pc), also presented in Fig.~\ref{fig:ScenarioAFormed}.  All MWBs found at
10~Myr in the N simulations {\em must} have formed dynamically.

What is most obvious is that the efficiency of MWB formation 
strongly depends on the density.  This should be of no
surprise as the formation rate depends on $n^3$.

Each of the scenarios are very similar, with 60--90 per cent of dense
simulations ($R_{1/2}=0.25$ and 0.5~pc) forming MWBs, but only 0--20
per cent of low-density ($R_{1/2}=1$ and 1.5~pc) simulations forming
MWBs (almost none at $R_{1/2}=1.5$~pc).

In Scenario N1 (blue solid line in Fig.~\ref{fig:ScenarioAFormed}), in which there are only two `massive' stars (all other
stars are $1 M_\odot$) a MWB
forms in the majority (70--80 out-of-100) of simulations at low 
$R_{1/2}$.  One of the reasons that the formation rate is so high when
there are only two stars that could form a MWB is that these stars
dynamically mass segregate, bringing them close together (increasing
$n^3$, and also increasing $1/\sigma^9$). 

Scenario N2 (green dashed line in Fig.~\ref{fig:ScenarioAFormed}) has two stars with masses greater than 10 M$_{\odot}$, 
and a range of low- and
intermediate-mass neighbours.  Only two thirds of the simulation contain a massive
wide binary at 10 Myr (less than in scenario N1).  This is not because
MWB have not formed, but due to the fact that once formed, a reasonable
fraction have been hardened by interactions with other stars, so that
their binary separation is less than 100 AU.  There therefore exists in
some of these simulations a population of massive, `tight' binaries
with separations $<100$~AU which we do not classify as MWBs (although these are {\em nowhere} near as tight as the few-day period massive star binaries commonly found in spectroscopic surveys).

In Scenario N3 (red dot-dashed line in Fig.~\ref{fig:ScenarioAFormed}), there are three stars with masses greater than 10 M$_{\odot}$, and a range of lower-to-intermediate-mass stars.  The number of simulations which form a massive wide
binary at small $R_{1/2}$ is slightly higher than in Scenario N1
(although note that the $\sqrt{N}$ `noise' on these numbers are about
$\pm 10$).  In this case the third massive star carries enough energy
to disrupt any newly formed MWBs and so these clusters are constantly
forming, then destroying, then forming etc. MWBs (cf. \citealt{Moeckel2011}). 

In Scenario N4 (cyan dotted line in Fig.~\ref{fig:ScenarioAFormed}), there are five stars with masses greater than 10 M$_{\odot}$ and a range
of lower-mass stars.  The situation is almost exactly the same as in
scenario N3 with a constantly forming and then destroyed population of
MWBs.  

In scenarios N2-N4, it is possible to have two MWBs present (two
pairs of the 5-20 available stars above 5 M$_{\odot}$), but this is rare and short-lived.

\bigskip

In summary, if no MWB is present at the start of a simulation then in
dense environments then {\em one} MWB is likely to form.  In low-density
environments it is very unlikely that a MWB will form.  

\subsection{The destruction and formation of MWBs}
\label{section: resultsScenarioB}
\begin{table}
\centering
\begin{tabular}{ | l | l | l | l | l |}
  \hline
  Scenario & \multicolumn{4}{| c |}{Number of Simulations in which the Original} \\
           & \multicolumn{4}{| c |}{Massive, Wide Binary Survived to $t=10$ Myr} \\
  \hline
     & $R_{1/2}=0.25$ pc & 0.50 pc & 1.00 pc & 1.50 pc \\
  \hline
  B1 & 100               & 100     & 100      & 100      \\     
  B2 & 68                & 72      & 92       & 97       \\     
  B3 & 11                & 22      & 72       & 89       \\     
  B4 & 7                 & 18      & 52       & 74       \\     
  \hline
\end{tabular}
\caption{Number of primordial WMBs which survived for 10~Myr, in clusters with different initial half-mass radii $R_{1/2}$, for each of Scenarios B1-B4.}
\label{table:scenarioBSurvived}
\end{table}

\begin{figure}
\centering
\includegraphics[]{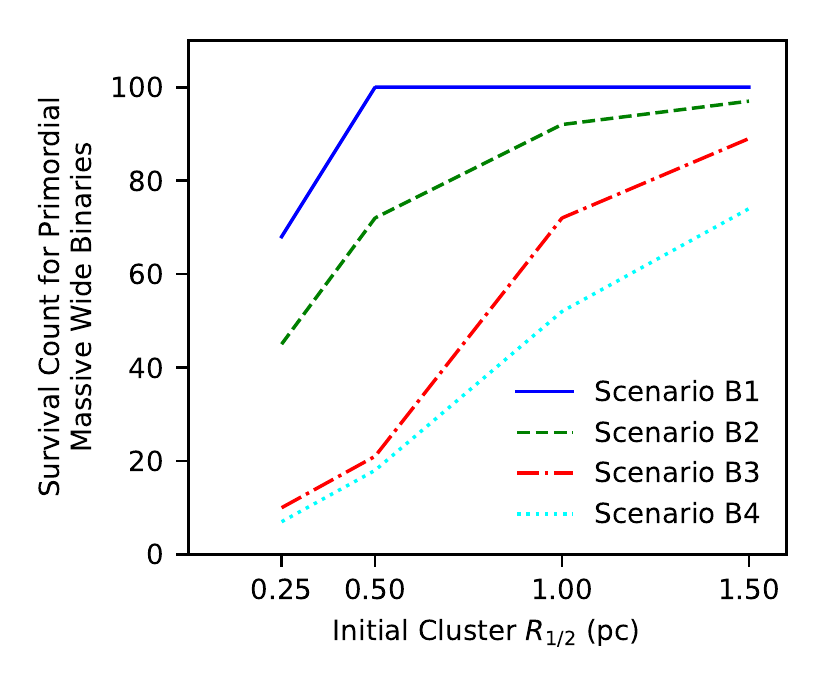}
\caption{Number, out of 100, of clusters with a primordial MWB in which the
  primordial MWB survives for 10~Myr, as a function
  of the initial cluster half-mass radius $R_{1/2}$, for Scenarios B1 (blue solid line), B2 (green dashed line), B3 (red dot-dashed line), and B4 (cyan dotted line).  \label{fig:ScenarioBSurvived}} 
\end{figure}

In Scenarios B1 to B4 all clusters have a primordial MWB.  But as we
have seen MWBs can form dynamically, and so in scenarios B3 and B4 it is quite
possible to have a MWB that is comprised of different stars to the
primordial MWB.  Therefore we distinguish between the survival of the
primordial MWB, and the presence of {\em any} MWB after 10~Myr (this may be the
primordial MWB, or may be a `new' MWB).

Table~\ref{table:scenarioBSurvived} gives the numbers (out-of-100) of
surviving {\em primordial} MWBs for scenarios B1-B4 for each density, this is 
illustrated in Fig.~\ref{fig:ScenarioBSurvived}.  

In Scenario B1 (blue solid line in Fig.~\ref{fig:ScenarioBSurvived}) there
are two massive stars in a MWB, and all of
the cluster stars are 1~M$_{\odot}$.  Here {\em all} of the primordial
MWBs survive regardless of density as the low-mass stars do not have
enough energy to disrupt the massive wide binary, but encounters do
harden around a quarter of the MWBs in the densest clusters ($R_{1/2}
= 0.25$) below our nominal MWB limit, hence the MWB fraction declines.

In Scenario B2 (green dashed line), the primordial MWB is surrounded by
other low-to-intermediate-mass stars.  At high densities
($R_{1/2} = 0.25$ and 0.5~pc) encounters can again harden a binary
below our MWB definition\footnote{To add a further complication, 
it is possible to destroy the
  primordial MWB, and then it reforms (cf. scenario B2), and then it can be 
  hardened below our MWB limit.}.  Therefore in around a third of systems with an primordial
MWB one is not present after 10~Myr, although this does depend on our
(somewhat arbitrary) definition of a MWB.

Scenarios B3 and B4 both have an primordial MWB, plus one or three 
(single) more massive stars. At high densities ($R_{1/2} = 0.25$ and
0.5~pc) the primordial MWB is very unlikely to survive.  In most cases
this is not because it is hardened below our definition, but rather
that it is destroyed by an encounter.  At lower densities ($R_{1/2} = 1$ and
1.5~pc) the survival of the primordial MWB is a matter of `luck' as to
whether it encounters the/one of the other massive stars in the cluster
or not, but 50--80/100 of the primordial MWBs are able to survive for 10~Myr (see the red and cyan lines).

\begin{figure}
\centering
\includegraphics[]{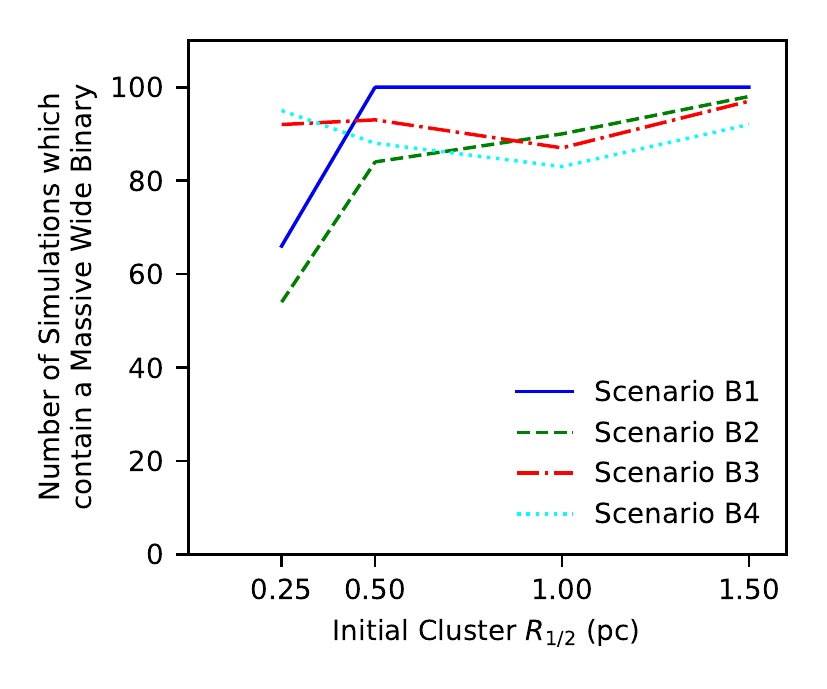}
\caption{Number, out of 100, of clusters which contained a primordial MWB which have 
  at least one MWB at 10~Myr, as a
  function of the initial cluster half-mass radius $R_{1/2}$, for Scenarios B1 (blue solid line), B2 (green dashed line), B3 (red dot-dashed line), and B4 (cyan dotted line).  
  \label{fig:ScenarioBTotal}} 
\end{figure}

In Fig.~\ref{fig:ScenarioBSurvived} we saw the fraction of primordial MWBs that survived.  However, as we saw in section 3.1, MWBs can be formed as well as destroyed.

In Fig.~\ref{fig:ScenarioBTotal} we show the number of simulations which
contain {\em any} MWB at 10~Myr, as a function
of the initial cluster half-mass radius $R_{1/2}$. 

In scenario B1 (blue solid line), any MWB must be the primordial  MWB (as there are only two stars capable of making-up a MWB), and so for B1 figs.~\ref{fig:ScenarioBSurvived} and \ref{fig:ScenarioBTotal} are identical. The reason that they are not 100\% at all densities is because a some of the surviving MWBs have been hardened below our nominal limit for a WMB, as explained above. 

This hardening effect also occurs in scenario B2 (green dashed line) where hardening is slightly more effective due to the presence of some stars $>1 M_\odot$).  The number of clusters with any MWB (fig.~\ref{fig:ScenarioBTotal}) is slightly higher than the numbers of primordial MWBs because other $\sim 5 M_\odot$ stars are present in the masses drawn from the IMF that can swap into the MWB, but this is a minor effect.

In scenarios B3 and B4 (red and cyan lines) there are one or three other massive stars, and some (typically about 8) $\sim 5 M_\odot$ stars are present in the masses drawn from the IMF.  Any of these other stars could pair to form a MWB.  In
fig.~\ref{fig:ScenarioBSurvived} we see that the primordial MWB rarely
survives at higher densities ($R_{1/2} = 0.25$ and 0.5~pc), but
fig.~\ref{fig:ScenarioBTotal} shows that the vast majority of these
clusters do contain a MWB at these densities: this is a `new' MWB
formed dynamically (as seen in Fig.~\ref{fig:ScenarioAFormed} where there are no
primordial MWBs).

\begin{figure}
\centering
\includegraphics[]{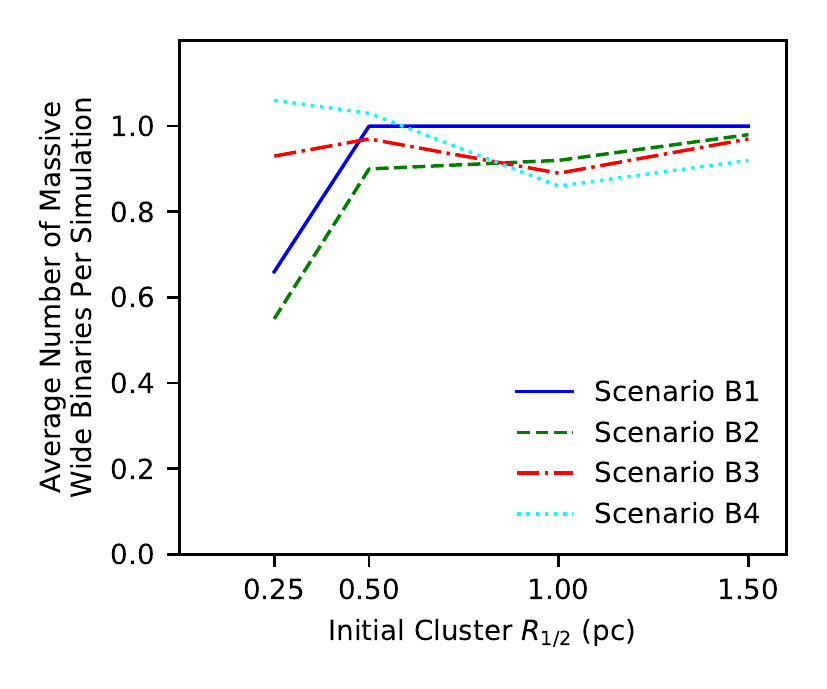}
\caption{Mean number of MWBs in each
  cluster that contained a primordial MWB after 10~Myr, as a function of the initial cluster
  half-mass radius $R_{1/2}$.  Error bars are $\pm 1 \sigma$ over the ensembles of 100 simulations.  For Scenarios B1 (blue solid line), B2 (green dashed line), B3 (red dot-dashed line), and B4 (cyan dotted line).
  \label{fig:ScenarioBAverage}} 
\end{figure}

\bigskip

In scenarios B2, B3 and B4 it is possible to have two MWBs present; this is rare, but does sometimes
happen.
Figure \ref{fig:ScenarioBAverage} shows the {\em mean} number of massive wide binaries found in each simulation at $t=10$ Myr, as a function of the initial cluster half-mass radius, for each scenario B1-B4 (i.e. for each different mass distribution).

Figure \ref{fig:ScenarioBAverage} shows that the expected number of MWBs in each cluster, given that each cluster initially contains one primordial MWB, is  close to unity.  The only times the number of MWBs is not about unity is Scenarios B1 and B2 at very high density ($R_{1/2} = 0.25$~pc), when binary hardening decreases the number of MWBs to an average of $\sim 0.6$ per cluster. 

In Scenarios B2, B3 and B4 there are usually about 10 stars that could potentially pair to make a MWB. However, due to the disruption of MWBs by other high-mass stars, only the most massive of MWBs will be  stable for a significant time.

\bigskip

\begin{figure}
\centering
\includegraphics[]{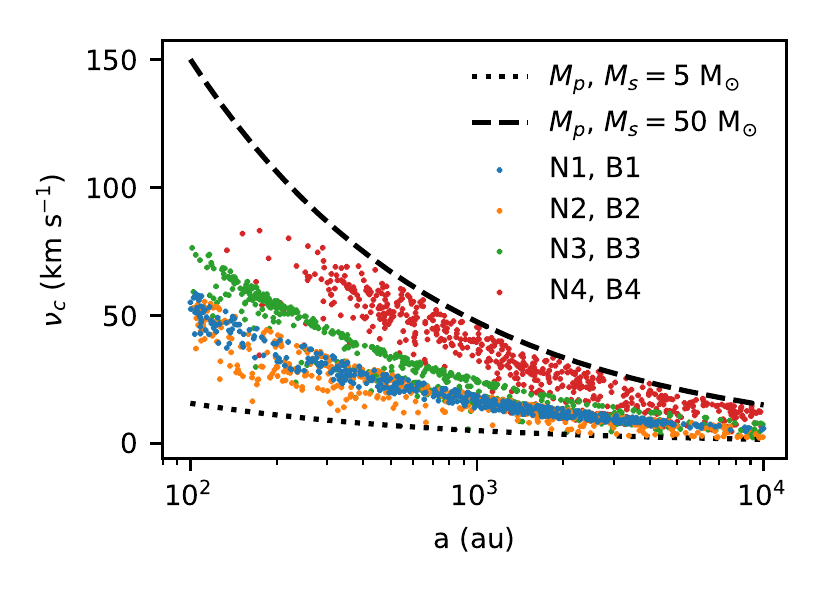}
\caption{The critical velocity, as defined in Eqn \ref{equation: criticalvelocity}, for each of the MWBs that were present at the end of each simulation for scenarios N1, B1 to N4, B4 assuming a perturber of mass $m_{\rm \rm pert}=1$ M$_{\odot}$. The dotted lines represent primary and secondary masses of 5 M$_{\odot}$ and the dashed lines represent primary and secondary masses of 50 M$_{\odot}$.}
  \label{fig:criticalvel1msun} 
\end{figure}

To help understand the survival of the MWBs, Fig. \ref{fig:criticalvel1msun} shows the critical velocity for destruction, as defined by Eqn \ref{equation: criticalvelocity}, for each of the MWBs that were present at the {\em end} of each simulation for all of our scenarios (N1--4 and B1--4) assuming a perturber mass of  
$m_{\rm pert}=1$ M$_{\odot}$.  The dotted line shows the critical velocity for the lowest possible mass MWB (5+5M$_{\odot}$), and the dashed line for the highest possible MWB mass 
(50+50M$_{\odot}$), for separations between $10^2$ and $10^4$~AU.

In fig.~\ref{fig:criticalvel1msun} all of the MWBs marked by coloured points (different colours for different scenarios) lie above the lower dotted line which is the critical velocity a 1 M$_{\odot}$ star must have to destroy the lowest-possible mass MWB (5+5M$_{\odot}$). This is exactly as expected as all simulations contain significant numbers of 1 M$_{\odot}$ stars and so should be hard enough to avoid destruction by these stars (although a soft binary could exist for a short time before being destroyed, see \citet{Moeckel2011}).

At any particular separation in Fig. \ref{fig:criticalvel1msun} increasing critical velocities for destruction mean increasing system masses (if $a$ is the same, then $m_{\rm p}$ and/or $m_{\rm s}$ must be greater for $v_{\rm c}$ to be larger).  

In Fig. \ref{fig:criticalvel1msun} the critical destruction velocities for MWBs in scenarios N1/B1 (blue points) lie in a fairly tight band as they are all $m_{\rm p} = 20$ M$_\odot$ and $m_{\rm s} = 10$--20 M$_\odot$ MWBs.  These MWBs are all well above the typical velocities of the 1 M$_{\odot}$ stars making-up the rest of the cluster and so they survive (although can be hardened below 100 au).

The critical destruction velocities for MWBs in scenarios N2/B2 (orange points) are more widely spread and to lower critical velocities than scenarios N1/B1 as some MWBs can form with a 5 M$_\odot$ companion from the cluster.

In scenarios N3/B3 (green points) almost all binaries are the 20 M$_\odot$ primary from the primordial MWB in a new MWB with the 30--35 M$_\odot$ `other' massive star leading to almost the same critical velocities at each separation.  The shift between the blue N1/B1 points and the green N3/B3 points is thus showing the difference in the masses of the two most massive stars that will pair up as a MWB.

Scenarios N4/B4 (red points) have five massive stars (possibly as high a mass as 50 M$_\odot$) and the spread represents whatever the two highest masses happen to be.

Hence the `hardness' of a system is much more representative of the masses available to combine into a MWB than the destructiveness of the environment.  The two most massive stars will pair into a wide binary which will almost certainly be hard enough to avoid destruction.

\subsection{Summary}

To quickly summarise the results we refer to $R_{1/2} = 0.25$ and
0.5~pc as `high-density' and $R_{1/2} = 1$ and 1.5~pc as `low-density'.

\noindent A) If no MWBs are present in a cluster they will very often form
dynamically at high-density, but not at low-density (see fig.~\ref{fig:ScenarioAFormed}).\\
B) Primordial MWBs will usually survive at low-density, and only be
destroyed at high-density if other massive stars are present (see
fig.~\ref{fig:ScenarioBSurvived}).\\
C) When primordial MWBs are destroyed at high-density they are usually
`replaced' by a new MWB (because of (A), see
fig.~\ref{fig:ScenarioBTotal}).\\
D) On average, one MWB will be found in each dense region (see
fig.~\ref{fig:ScenarioBAverage}).

{\em The only environment we simulate in which we do not usually
see just a single MWB present at 10~Myr are low-density clusters that did not have
a primordial MWB.}

\section{Discussion}
\label{section: discussion}

In most environments we simulated, almost always a {\em single} MWB is present at 10~Myr.  The only environments in which MWBs are rare
are low-density environments which never had a MWB.  

This is because of two competing effects:\\
MWBs are `hard' to lower-mass stars (which do not
carry enough energy to disrupt the MWB), but `soft' or `intermediate'
to other massive stars (which do carry enough energy).  Therefore they
are destroyed when other massive stars are present in an environment
dense enough to allow encounters.\\
MWBs readily form in dense environments due to the $m^{5}n^{3}$
dependence of the binary formation rate. (The massive star density in
dense clusters is also enhanced by rapid dynamical mass segregation
increasing $n^3$ significantly).

An important point is that if a MWB is present there is almost always
only a single MWB.  MWBs are soft/intermediate in the presence of other massive
stars which means they are constantly being destroyed and formed when
other massive stars are present \citep{Moeckel2011}.
The balance between the formation and destruction of MWBs in dense environments means that they are a probe of the past density history of a region as we show below for Cyg~OB2. 

Very usefully observationally, a MWB has two massive (ie. bright), widely-separated, components
means that they should be observable in fairly distant regions (at
least a few kpc) where the low-mass population is much more difficult to observe.

\subsection{The past history of Cyg~OB2}

Cyg~OB2 is a $2-10$~Myr old, $\sim 10^5 M_\odot$ unbound association with a current size of $\sim 20$~pc, with a 3-dimensional velocity dispersion of $\sim 18$ km s$^{-1}$ Cyg OB2 is unbound (\citealt{Wright2016} and references therein).  That Cyg OB2 is {\em currently} unbound makes determining its past dynamical history difficult.  It is possible that it was one, or several, initially bound (sub)clusters that have each become unbound (due to gas expulsion?), or that it was always globally unbound.  We argue that the MWB population is a useful tracer of the past density history.

Caballero-Nieves et al. (in prep., hereafter CNip) have observed a sample of 74 O-star primaries in Cyg~OB2 to search for wide 100--10000~AU companions (it is somewhat more subtle than this as detection depends on separation and magnitude difference).  CNip are able to detect more distant companions to a mass of (very roughly) $4 M_\odot$ at wider separations (hence our adoption of 5~M$_\odot$ as a `massive' star).

What is important for our discussion here is that CNip find a wide, massive companion for 38 of the 74 primaries ($\sim 51$ per cent MWB fraction).  Note that it may well
be that one or both components of each MWB are themselves close
binaries -- this makes no difference to our argument.

There are three ways in which we can explain the large number of MWBs in
Cyg~OB2.  

Firstly, that many massive stars in Cyg~OB2 formed in low-density
environments in primordial MWBs.  Therefore what we observe are a
large number of primordial MWBs.

Secondly, that massive stars formed in {\em many} small, dense groups
(either in primordial MWBs or not), and each group formed (on average)
about one MWB.  Therefore what we observe are a large number (at least 40) of dynamically formed MWBs, roughly one per sub-region.

Thirdly, some mixture of the first and second possibilities, with the observed population being a mix of primordial and dynamically-formed MWBs.

Whichever of the three possibilities is correct it means that {\em massive
star formation in Cyg~OB2 was widely distributed}.  It was either
almost completely isolated, or in many small, dense groups (or some
mix of these): but it could not have been as a single (or even a few)
massive `clusters'.  

This is in agreement with \citet{Wright2014} and
\citet{Wright2016} who argue from the distribution and kinematics of
Cyg~OB2 that it has always been widely distributed and unbound.

Cyg~OB2 has a standard IMF, i.e. has the number of massive stars expected for a region of $10^5 M_\odot$ (\citealt{Wright2015}).  The number of MWBs very strongly suggests that there were many sites of massive star formation that did not know about each other (they never interacted dynamically, otherwise we would not see so many MWBs).  Therefore, whatever mechanism forms massive stars must be able to `randomly sample' the IMF, e.g. it can form very massive stars (up-to $100 M_\odot$ in Cyg~OB2) without `knowing' that the total mass of the region is very large.  This argues strongly against `deterministic' models for the origin of massive stars, e.g. the classic version of competitive accretion, \citep{Bonnell1997}, and suggests the cluster mass-maximum stellar mass relationship is statistical rather than fundamental (\citealt{Weidner2004}; \citealt{Parker2007}).

\subsection{How to use the numbers of MWBs}

More generally, in any region one can think of four possibilities in terms
of the numbers of MWBs that are present:\\
1) Currently high-density with very few or no MWBs.  No information on the
primordial MWB population as most/all would have been destroyed if
they existed.  The region could have been lower density in the past and
collapsed, or always high density.\\
2) Currently low-density with very few or no MWBs.  If the region was denser in the past
that would have destroyed most/all primordial MWBs, if it was always
low-density then there were few/no primordial MWBs.\\
3) Currently high-density with many MWBs.  This is unexpected: it must have spent only a
little time at a high-density otherwise we would expect all but one (or
two) primordial MWBs to have been destroyed, and no more than one (or
two) to possibly have formed.\\
4) Currently low-density with many MWBs (e.g. Cyg~OB2).  Either the region was always
low-density with many primordial MWBs, or it contained many small `sub-clusters' that could each form a MWB. 

Our wording has been rather woolly here in terms of `high-density/low-density' or
`number of MWBs'.  How many MWBs
are significant depends on the number of massive stars that are
present to pair into MWBs, and the masses of those stars relative to those around it.  It is difficult to say much from only two
massive stars either being in a MWB or not.  However, apparently half
of the large population of massive stars in Cyg~OB2 being in MWBs is
clearly significant (`many').  The point at which `many' becomes `few'
is less clear, and is a judgement call based on the details of any particular region
that is being examined.

\section{Conclusion}
\label{section: conclusion}
We define Massive Wide Binaries (MWBs) as binary systems containing
two stars of mass $> 5 M_\odot$ with separations between $10^2$ and
$10^4$~AU (ie. bright, visual binaries in the high-mass tail of the IMF).

We examine the interplay between the destruction and formation of MWBs
in (virialised Plummer sphere) clusters of total mass $\sim 600
M_\odot$ ($\sim 400$ stellar members) using $N$-body simulations.

Our clusters always either have a
`primordial' MWB or just two single massive stars.  The rest of the
stars in the cluster are: (a) all Solar-mass; (b) an IMF with no other
 stars more massive than $10 M_\odot$; (c) an IMF with one other (more) massive star; or (d)
an IMF with three other (more) massive stars.  For each mass range we
run ensembles of 100 simulations for 10~Myr with half-mass cluster
radii of 0.25, 0.5, 1 and 1.5~pc. 

Our main results can be summarised as follows:\\
1) Primordial MWBs almost always survive in low-density environments, or
any environment with no other massive stars;\\
2) Primordial MWBs are usually destroyed in high-density environments
when other massive stars are present;\\
3) A single MWB very often forms dynamically in high-density environments;\\
4) MWBs rarely form dynamically low-density environments.

The combination of these results means that the only (local) environment in
which no MWB will be present is a low-density
cluster which contained no primordial MWB.  In all other (local) environments
either a single primordial MWB will survive, or (almost always) a single MWBs can be formed dynamically.

Therefore, any region containing many MWBs must have either be (or 
have been) many high-density sub-clusters (which form one MWB each),
many primordial MWBs which never encountered another massive star, or
some mixture of both.  What it could not have been is a single, dense
cluster (or fewer dense (sub-)clusters than there are MWBs).

The low-density association Cyg~OB2 has approximately 40 MWBs (with a MWB
fraction of roughly a half).  This is further evidence that Cyg~OB2
has always been globally diffuse, and must have contained either many
(at least about 40) small high-density regions in which to either dynamically form
MWBs, or contained many primordial MWBs that have always been in
low-density environments.  That Cyg~OB2 as a whole has as many massive stars as would be expected for its total mass, suggests that massive star formation `randomly samples' the IMF (in that Cyg~OB2 `knew' to form very massive stars even though they knew nothing about each-other dynamically).




\bibliographystyle{mnras}
\bibliography{Wide_Binaries} 

\bsp	
\label{lastpage}
\end{document}